\documentclass[12pt]{article}
\usepackage{graphicx}
\begin{document}
\centerline{\large \bf Evolutionary ecology in-silico: }
\centerline{\large \bf evolving foodwebs, migrating population and speciation}

\medskip
\centerline{Dietrich Stauffer{\footnote{Visiting from Institute for Theoretical Physics, Cologne University, D-50923 K\"oln, Euroland; e-mail:stauffer@thp.uni-koeln.de}}, Ambarish Kunwar{\footnote{ e-mail: ambarish@iitk.ac.in}} and Debashish Chowdhury{\footnote{Corresponding author; e-mail: debch@iitk.ac.in}}}

\medskip
\noindent
\centerline{Department of Physics}
\centerline{Indian Institute of Technology}
\centerline{Kanpur 208016} 
\centerline{India}

\medskip
\noindent

\medskip
\noindent

\vspace{4cm}

\noindent {\bf Abstract:}
We have generalized our ``unified'' model of evolutionary ecology by
taking into account the possible movements of the organisms from one
``patch'' to another within the same eco-system. We model the spatial
extension of the eco-system (i.e., the geography) by a square lattice
where each site corresponds to a distinct ``patch''. A self-organizing
hierarchical food web describes the prey-predator relations in the
eco-system. The same species at different patches have identical food
habits but differ from each other in their reproductive characteristic
features. By carrying out computer simulations up to $10^9$ time steps, 
we found that, depending on the values of the set of parameters, the 
distribution of the lifetimes of the species can be either exponential 
or a combination of power laws. Some of the other features of our 
``unified'' model turn out to be robust against migration of the 
organisms.


\newpage

\section{Introduction}

Direct experimental evidence in recent years
\cite{thompson98a,thompson98b,thompson99,stockwell03,turchin03,
yoshida03,fussmann03} has established that significant evolutionary
changes can occur in an ecosystem over ecologically relevant time 
scales. In other words, the dynamics of ecology and evolution are 
inseparable. Significant progress has been made over the last two 
years in developing detailed models that describe not only ecological 
phenomena on short periods of time but also evolutionary processes 
on longer time scales \cite{chow1,chow2,chow3,chow4,rikvold,hall,
collobiano,quince,droz}. Some of these models describe the population 
dynamics of each species in terms of only a single dynamical variable, 
whereas some other more detailed models monitor the birth, ageing 
and death of individual organisms. A thorough comparison of these 
recent models will be given elsewhere \cite{chow5}.

In all the versions of our ``unified'' model \cite{chow1,chow2,chow3,chow4} 
so far we have assumed that the population of the prey as 
well as that of predators are uniformly distributed in space. 
The absence of the spatial degrees of freedom in such models of 
spatially-extended systems should be interpreted, using the language 
of statistical physics \cite{chowstau}, as a mean-field-like 
approximation. This situation is similar to a well-stirred chemical 
reaction where the spatial fluctuations in the concentrations of 
the reactants and the products is negligibly small. On the other 
hand, spatial inhomogeneities in the eco-systems and migration of 
organisms from one eco-system to another are known to play crucial 
roles in evolutionary ecology \cite{czaran,bascompte,tilman}.

In recent years, the spatial inhomogeneity of the populations observed
in real ecosystems have been captured by extending the Lotka-Volterra
systems on discrete lattices where each of the lattice sites represents
different spatial ``patches'' or ``habitats'' of the ecosystem
\cite{tainaka89,satulovsky94,boccara94,frachebourg96,lipowski99,
antal01a,antal01b,droz02,johst99}.
However, almost all these investigations of ``geographical'' effects
were restricted to only a single pair of species one of which is the
predator while the other is the prey.

For modelling the population dynamics of more than two species, one
needs to know the {\it food web} which is a graphical way of
describing the prey-predator relations, i.e., which species eats
which one and which compete among themselves for the same food
resources \cite{pimm,polis,drossel03,cohen90}. More precisely, a
food web is a directed graph where each node is labelled by a
species' name and each directed link indicates the direction of
flow of nutrient (i.e., {\it from} a prey {\it to} one of its
predators).

A static (time-independent) food web may be a good
approximation over a short period of time. But, a more
realistic description, valid over longer periods of time, must
take into account not only the adaptations of the species and
their changing food habits, but also their extinction and
creation of new species through speciation \cite{gavrilets}. 
These processes make the food web a slowly evolving graph. 
Such slow time evolution of the food webs were naturally 
incorporated in our earlier mean-field descriptions 
\cite{chow1,chow2,chow3,chow4}.

In this paper we extend the latest version of our ``unified'' 
model \cite{chow4} of evolutionary ecology incorporating spatial 
inhomogeneities of the eco-system, i.e., spatial variations from 
one patch to another. Biologically motivated simulations on 
lattices, to take into account the geographical extent of an 
ecosystem, have a long tradition, e.g. for prey-predator relations 
\cite{pekalski}, ageing \cite{makowiec}, speciation \cite{sousa}; 
their results sometimes \cite{shnerb} differ drastically
compared with those of the homogeneous models that ignore the
spatial extension. Such type of simulations are part of the 
widespread efforts to apply statistical physics methods to biology 
\cite{drossel} or other fields outside physics \cite{moss}.

Our model \cite{chow1,chow2,chow3}, in its latest and most realistic 
form \cite{chow4}, allows each individual organism to give birth to 
$M$ offspring at each time step, provided the age of the adult 
organism is above some minimum reproduction age $X_{rep}$ (which 
varies from one species to another), with a probability that depends 
on the current age as well as $X_{rep}$. Similarly, each organism 
can die, for genetic reasons, with a probability (mortality rate) 
increasing exponentially with age, up to some species-dependent 
maximum age $X_{max}$. Moreover, mutations happen at each time step, 
with the probability $p_{mut}$, which change $X_{rep}$ and $M$ 
randomly by $\pm 1$.

In our model, we assumed a simple hierarchical food web where the
species are arranged in trophic levels $\ell$
($1 \le \ell \le \ell_{max}$), with no more than $m^{\ell - 1}$
species in each level. $\ell=1$ thus represents one species
at the top of all food chains; typically, $m=2$. Species in level
$\ell$ can prey on (or ignore) species on the immediately lower
level $\ell +1$. Moreover, the prey-predator relations are dynamic;
because of random mutation, the food habits of the species change.
Thus, an individual organism may fall victim to one of its predators.
However, scarcity of food may also lead to starvation deaths among
the predators.

A species becomes extinct when all the corresponding population
vanishes. We re-fill each ecological niche left empty by an extinct
species, separately with probability $p_{sp}$, by another species
on the same (or a lower) level; independent random mutation of the
parent and daughter species leads, eventually, to two different 
species.

If the total biomass of the surviving individuals in the whole
ecosystem is below the preassigned maximum possible value, a new
trophic level is created, with probability $p_{lev}$, i.e.
$\ell_{max}$ is increased by one. The actual number of occupied
levels can fluctuate with time depending on extinction and speciation.

The maximum possible age of individual organisms, $X_{max}$, for
each species is assumed to depend only on the trophic level and
fixed at $100/2^{(\ell-1)/2}$. The minimum reproduction age
$X_{rep}$ is initially distributed randomly between one and this
maximum age. We do not monitor the ageing and death of the the
``bacteria'' occupying the lowest level of the food web; instead,
we assume a constant population of the ``bacteria'' throughout
the evolution.

The paper is organized as follows. In section 2 we study a simplified 
version of our ``unified'' model that provides a coarse-grained 
description of evolutionary ecology where the birth, ageing and 
death of the individual organisms do not appear explicitly. 
After studying the homogeneous version of this model we also 
investigate the effects of inhomogeneities arising from population 
migration. A few variants of this model are considered. In section 
3 we extend the model formulating it in terms of the detailed 
``micro-dynamics'' of individual organisms which can also migrate 
from one ``patch'' of the eco-system to another. The main results 
are summarized and conclusions are drawn in section 4.

\section{A coarse-grained description of evolutionary ecology}

In this section we do not monitor the ``microscopic'' dynamics of
birth, ageing and death of the individuals explicitly. Instead,
we describe the population dynamics of each species, in a
coarse-grained manner, in terms of a single dynamical variable.
In the first subsection below we ignore the spatial fluctuations
while in the second subsection we take into account the fluctuations
from one ``patch'' to another by modelling the eco-system as a
lattice where each site represents a ``patch''.

\subsection{Homogeneous approximation}

\begin{figure}[hbt]
\begin{center}
\includegraphics[angle=-90,scale=0.33]{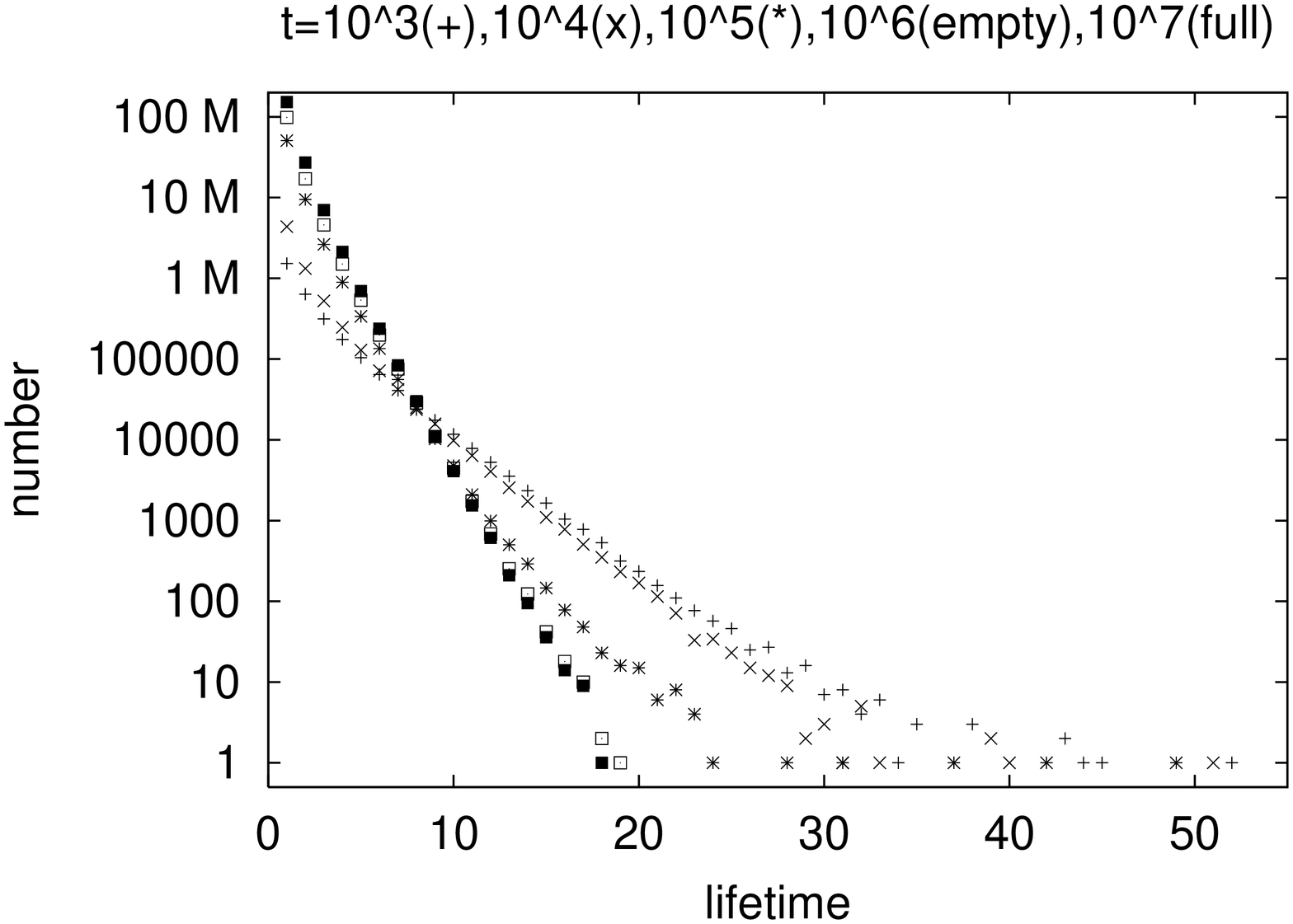}
\includegraphics[angle=-90,scale=0.33]{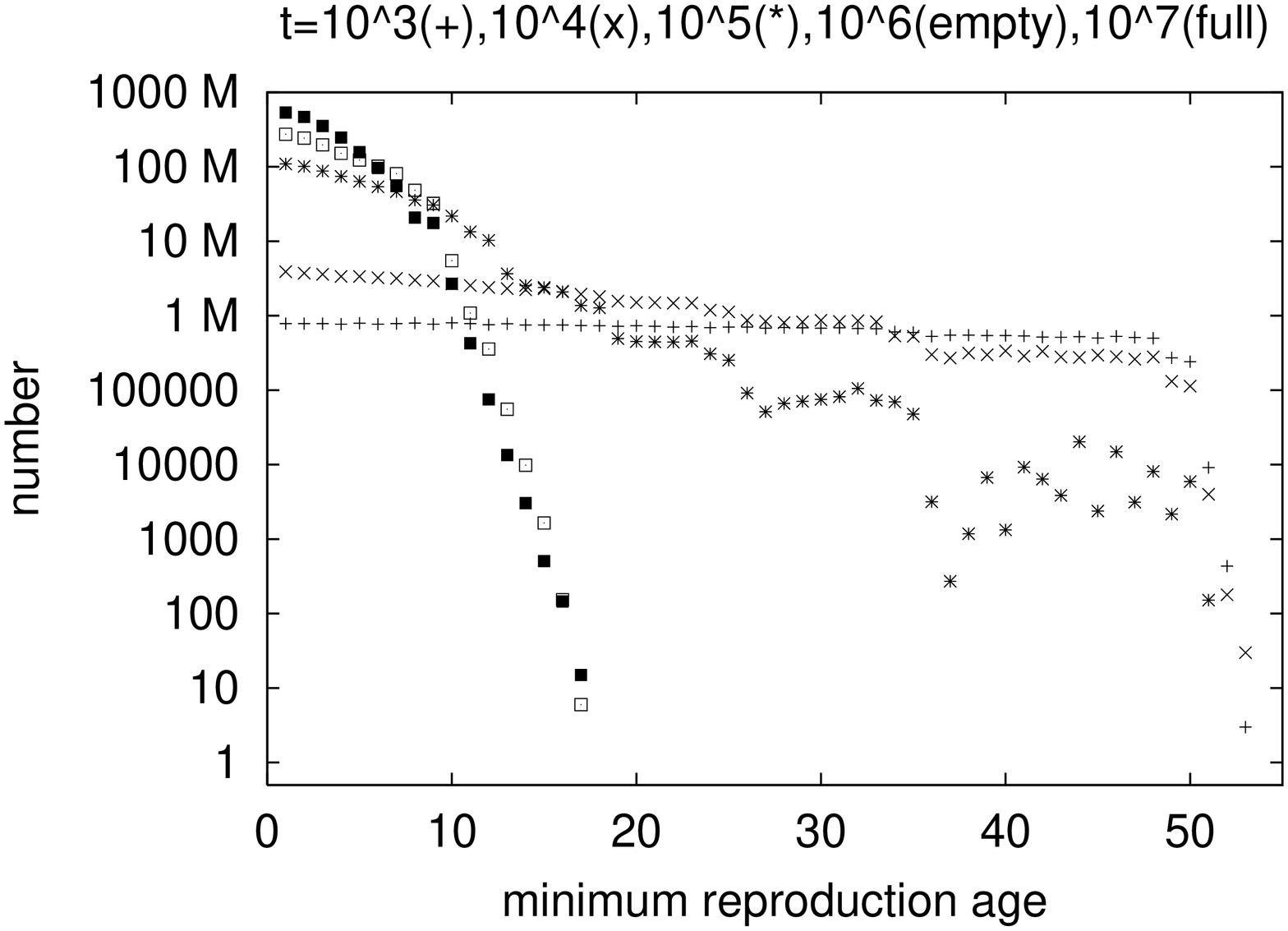}
\end{center}
\caption{Histogram for the lifetimes (part a) and minimum reproduction times
(part b) of species, without lattice, with
observation time $t = 10^3, 10^4, 10^5, 10^6, 10^7$ from right to left, summing
over $10^4, 10^3, 10^2, 10, 1$ samples, respectively.
}
\end{figure}

\begin{figure}[hbt]
\begin{center}
\includegraphics[angle=-90,scale=0.33]{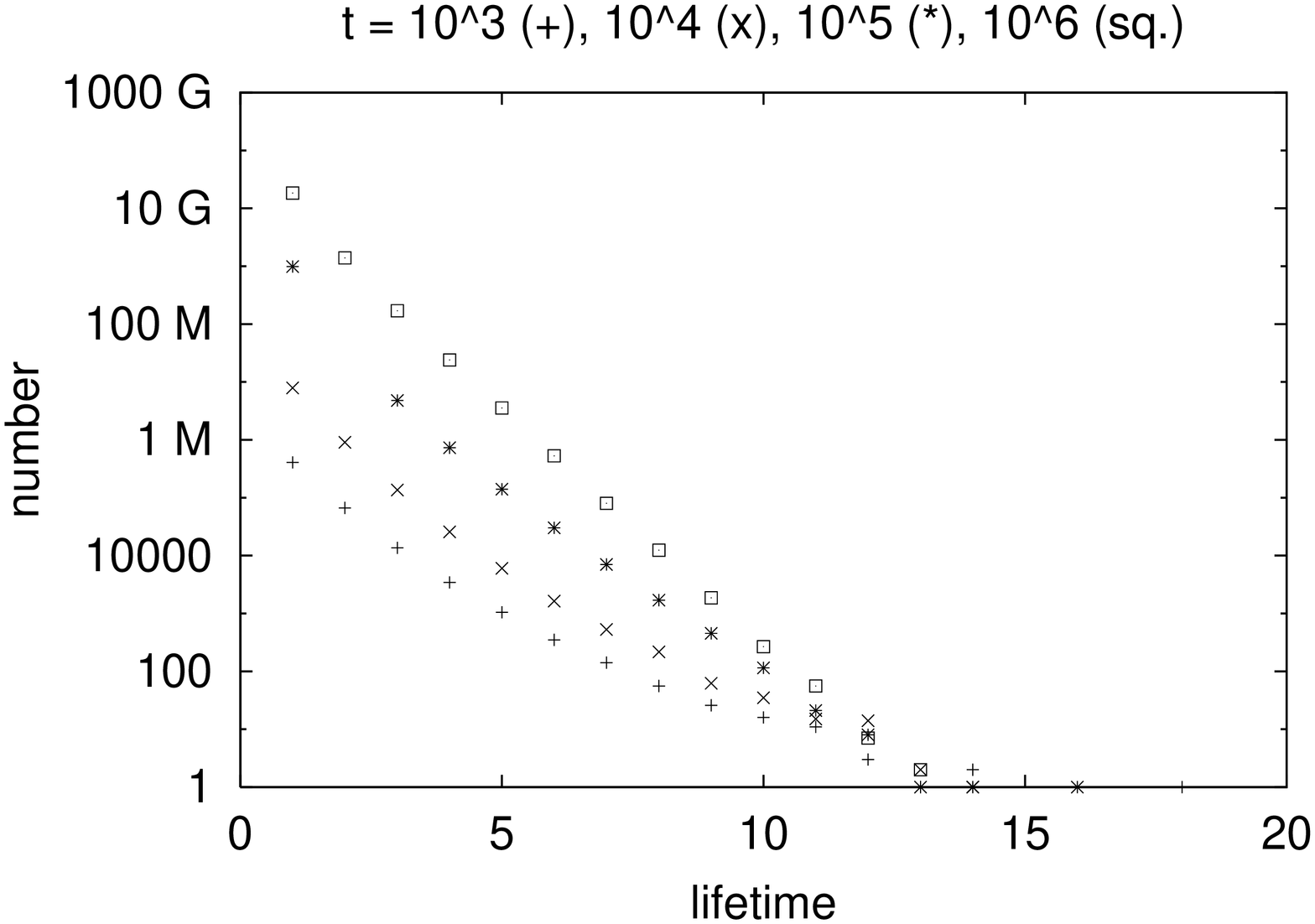}
\includegraphics[angle=-90,scale=0.33]{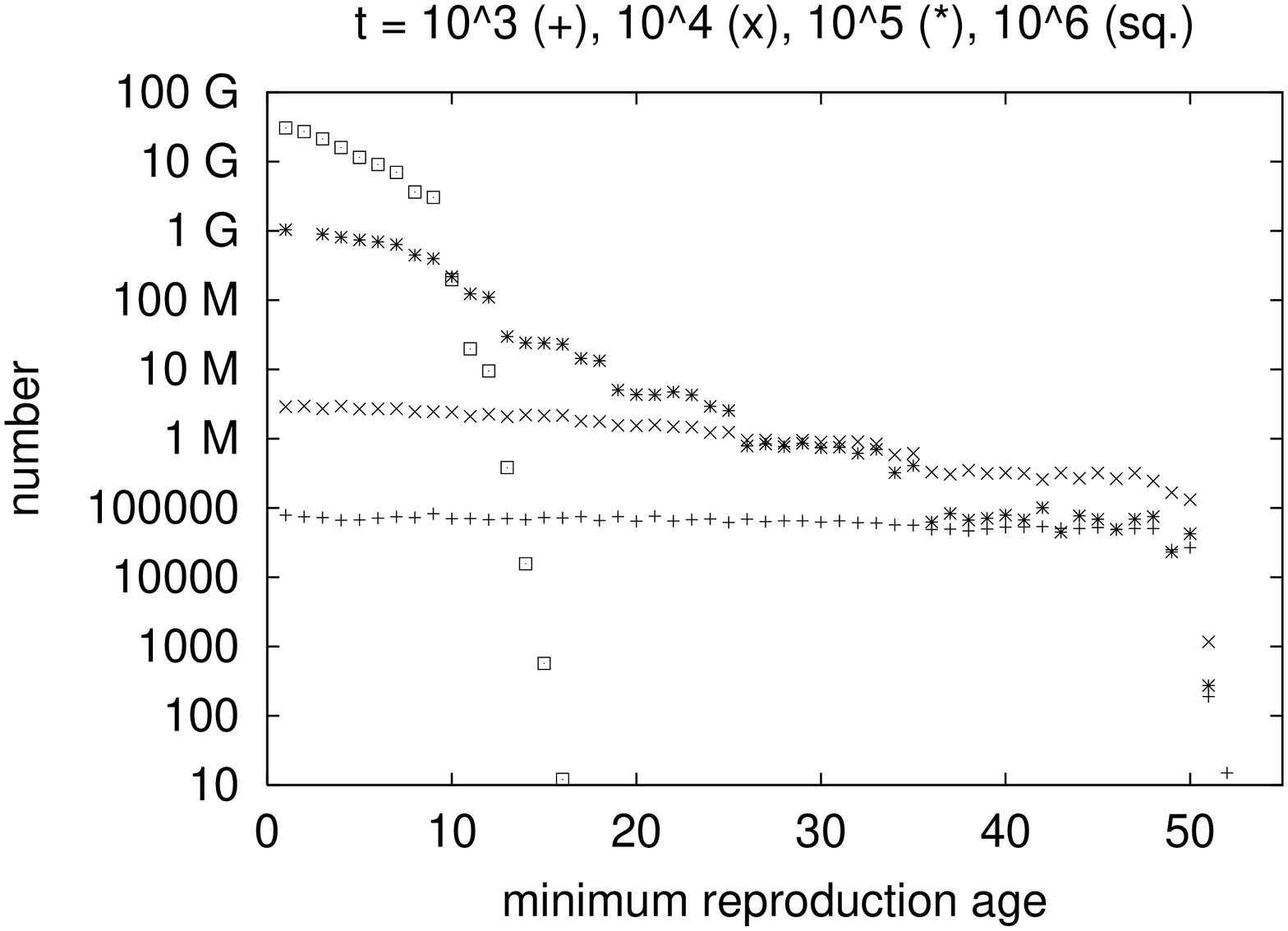}
\end{center}
\caption{Histogram for the lifetimes (part a) and minimum reproduction times
(part b) of species, with diffusion on a
$31 \times 31$ square lattice, with
observation time $t = 10^3, 10^4, 10^5$ from right to left, using
one sample only.
}
\end{figure}

\begin{figure}[hbt]
\begin{center}
\includegraphics[angle=-90,scale=0.5]{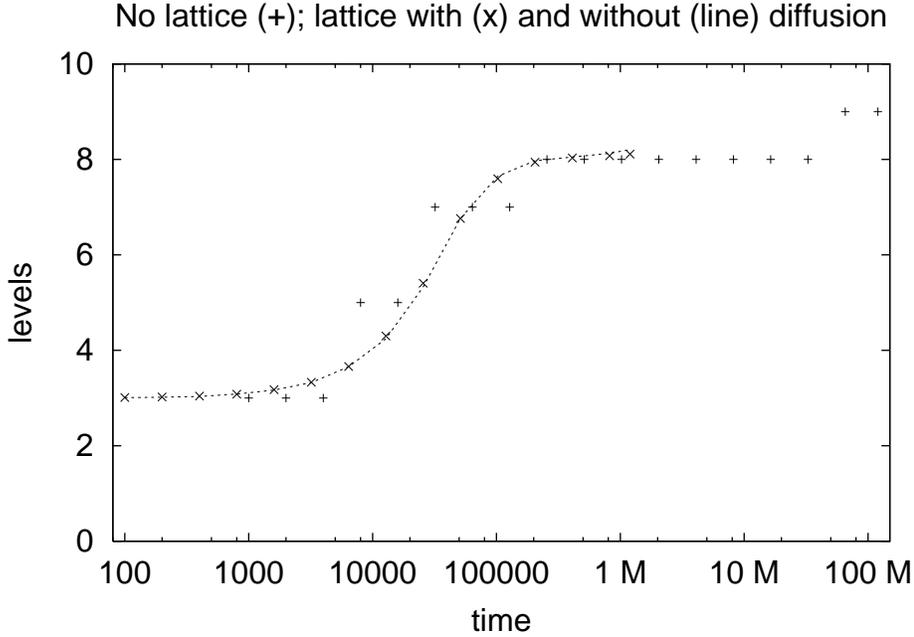}
\end{center}
\caption{
Variation with time for the average of the maximal number of levels, with
(x) and without (line) lattice (same simulations as in Figs.1
to 4.). The + symbols show one run without lattice to much longer times.
These data show the need to iterate up to $10^6$ time steps for  
equilibration.
}
\end{figure}

When we deal with species and no longer with individuals, we have to omit
the individual births and genetic deaths and replace them by a fitness of the
whole species; normalized to unity this fitness gives the survival probability
of the species as a whole. Our fitness has two components, both taken from
the birth and death probabilities of our previous model \cite{chow4}:

a) High birth rates $M$ decrease the life expectancy of the parent; thus
a species is assumed to survive a time step with probability exp($-rM)$,
where $r$ is taken as 0.05 as in \cite{chow4}.

b) Having survived step a, a species is the fitter, the higher the number
$M$ of offspring per time step is and the lower the minimum age of
reproduction $X_{rep}$ is. Thus each species survives with probability
$[(X_{max}-X_{rep})M]/[(X_{max}-X_{min})M_{max}]$ where $M_{max}$ is the
current maximum litter size $M$ in the entire ecosystem, and $X_{min}$ 
the current minimum value of the $X_{rep}$ in the whole ecosystem.
In this way, the fittest species, and only this
species, survives for sure, all others survive with the above probability only.

[Another modification is required in the prey-predator deaths. The Verhulst
deaths are now omitted, and instead of a reduction of the population by the
amount given in eq(1) of \cite{chow4}, the population goes extinct with
a probability given by this amount.]

We took $p_{sp}=0.1, \; p_{mut} = 0.001, \; p_{lev} = 0.0001, \; r = 0.05, \;
C = 0.2$ as in \cite{chow4}, and allowed the creation of a new level only if
at most four species were alive above the lowest level. We averaged over $t$
consecutive time steps,
after making $t/5$ time steps to establish a reasonable equilibrium.

We found that the number of food levels actually occupied at any one moment
(between 1 and 9) is normally distributed (``Gaussian'' parabola on a semi -
logarithmic plot; not shown). Thus for extremely long times $t$ the maximum
number of food levels should increase with the square-root of log($t$). This
extremely slow increase is compatible with one simulation (Figs.3 and 7) for
$ t = 10^8$. Biologically, longer times do not make much sense; 100 million
years ago we did not have the genus {\it homo} on earth.

The histograms for the lifetimes of species, and for the minimum
reproduction times $X_{rep}$, Fig.1, both show that $t = 10^3$ and $10^4$
are too short, $t = 10^5$ is intermediate between short- and long-time
behaviour, and $t = 10^6$ and $10^7$ give barely distinguishable long-time
behaviour. For this comparison we summed up $10^7/t$ samples so that the
total number of measurements is the same in all five cases.

Fig.1a also shows an exponential lifetime histogram for long $t$, as opposed to
the more complicated histogram in \cite{chow1,chow2,chow3,chow4} and the simple
power law often claimed for reality \cite{drossel,newman02}. Thus
the treatment of individuals, not the other complications, seems to be
responsible for the complex lifetime distributions in these earlier models.
Fig.1b, on the other hand, is not much different from Fig.3 in \cite{chow4}.
The average litter size $M$ fluctuates strongly, increasing overall perhaps
logarithmically with the number of time steps (not shown).

\subsection{Spatial inhomogeneity}

With a whole ecosystem of several food layers on every site of a lattice,
computational requirements increase drastically. To reduce memory requirements
we thus assume that the  prey-predator relations are the same on each lattice
site; only the other species-dependent properties like $X_{rep}$ and $M$, and
the number $\ell_{max}$ of food levels, can change from site to site.  Mutations
thus
act on the latter quantities differently for different sites, while mutations
on prey-predator relations act simultaneously on the whole lattice, for the
given niche (= possible species in ecosystem). If the expansion of one species
to another site, where it later undergoes different mutations, is called
speciation (i.e. parapatric speciation), then the previously discussed
speciation events which also change the prey-predator relations, are changes of
a higher taxonomical level, like changes of genus. Alternatively, if we
call the changes of the preceding section a speciation, the changes due to
expansion onto two lattice sites allows the formation of sub-species (``races'')
of the same species, sharing the same predator-prey relations.

If we only allow for different ecosystems on different lattices sites, the
results are roughly equivalent to averaging over many separate runs of a
single site, as we did in the previous section.  Only the fluctuations are
reduced due to the agreement for prey-predator relations. More interesting
and realistic is motion of species across the lattice. Thus at every time step 
each non-extinct species with probability 1/2 stays where it is, and otherwise
selects randomly one of its four neighbour sites on the square lattice (using
periodic boundary conditions). If that site is occupied at the corresponding
niche (i.e. at the same level and on the same node) then again nothing happens;
if the selected neighbour site is not already occupied for that niche, then
the old species occupies this place but also stays at its old site. Thus we
have an expansion, not a mere displacement, of the species. From then on, both
sites undergo the usual mutations and thus slowly the two populations drift
apart genetically to become different (sub-)species.

Apart from somewhat shorter lifetimes, our results  in Fig.2 look
similar to Figs. 1: Simple exponential decay for the
distributions of lifetimes and minimum reproduction ages, for long enough times.
More interesting is Fig.3 with the maximal number of levels, $\ell_{max}$,
averaged over all
lattice sites. As a function of time, this quantity at first increases roughly
linearly in time from its initial value 3. After about $10^5$ time steps,
a much slower increase is observed, leading to a plateau near 8.
Diffusion barely changes the lattice results without diffusion, and even
without a lattice the results are about the same except that the single run
without a lattice shows stepwise increase of $\ell_{max}(t)$.

\begin{figure}[hbt]
\begin{center}
\includegraphics[angle=-90,scale=0.33]{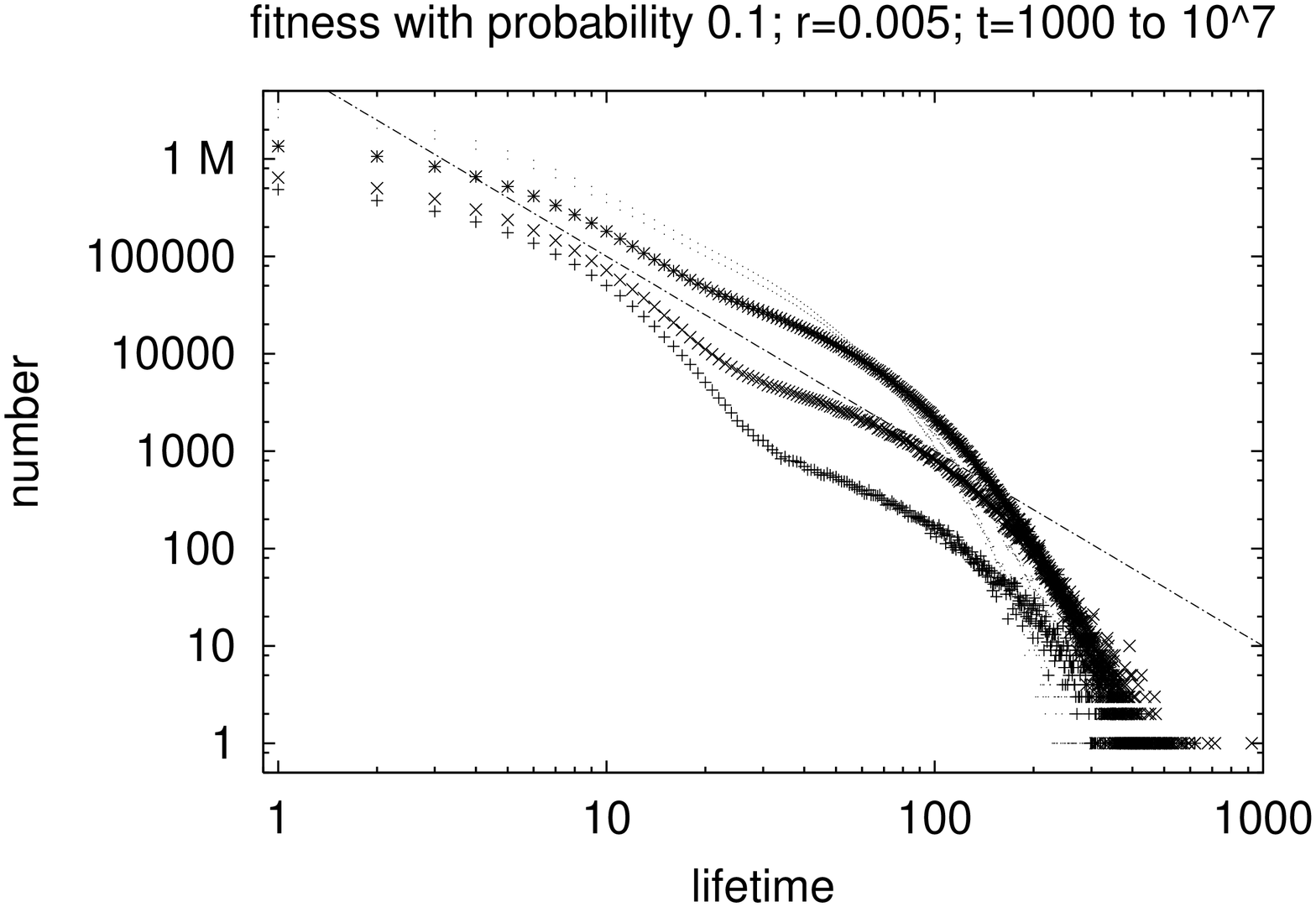}
\includegraphics[angle=-90,scale=0.33]{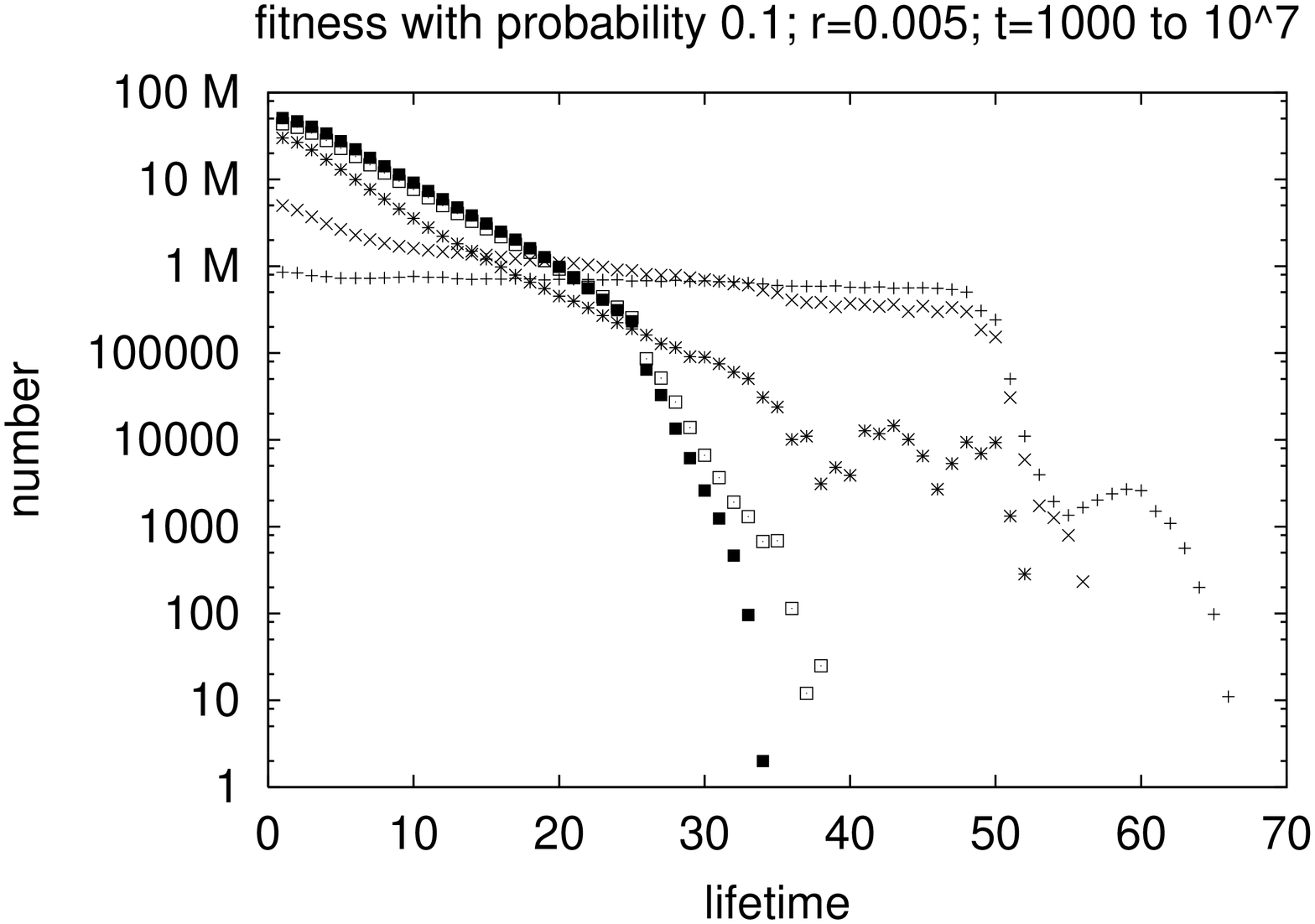}
\end{center}
\caption{As Fig.1, but with fitness tested every tenth time step, and $r$
ten times smaller. The two longest times are shown by small dots in the left
part. The straight line has the slope --2 sometimes claimed for real extinctions.
}
\end{figure}

\begin{figure}[hbt]
\begin{center}
\includegraphics[angle=-90,scale=0.33]{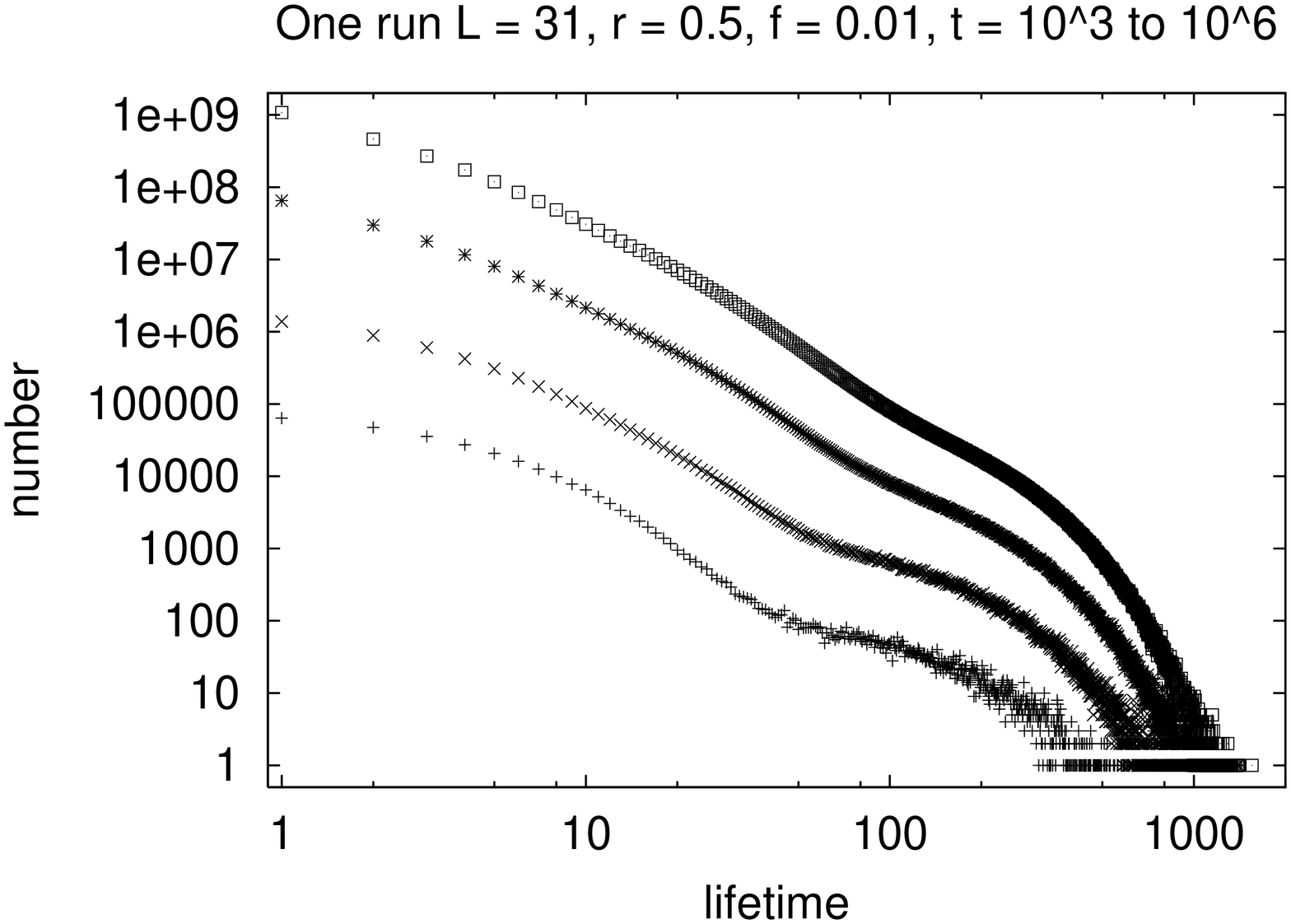}
\includegraphics[angle=-90,scale=0.33]{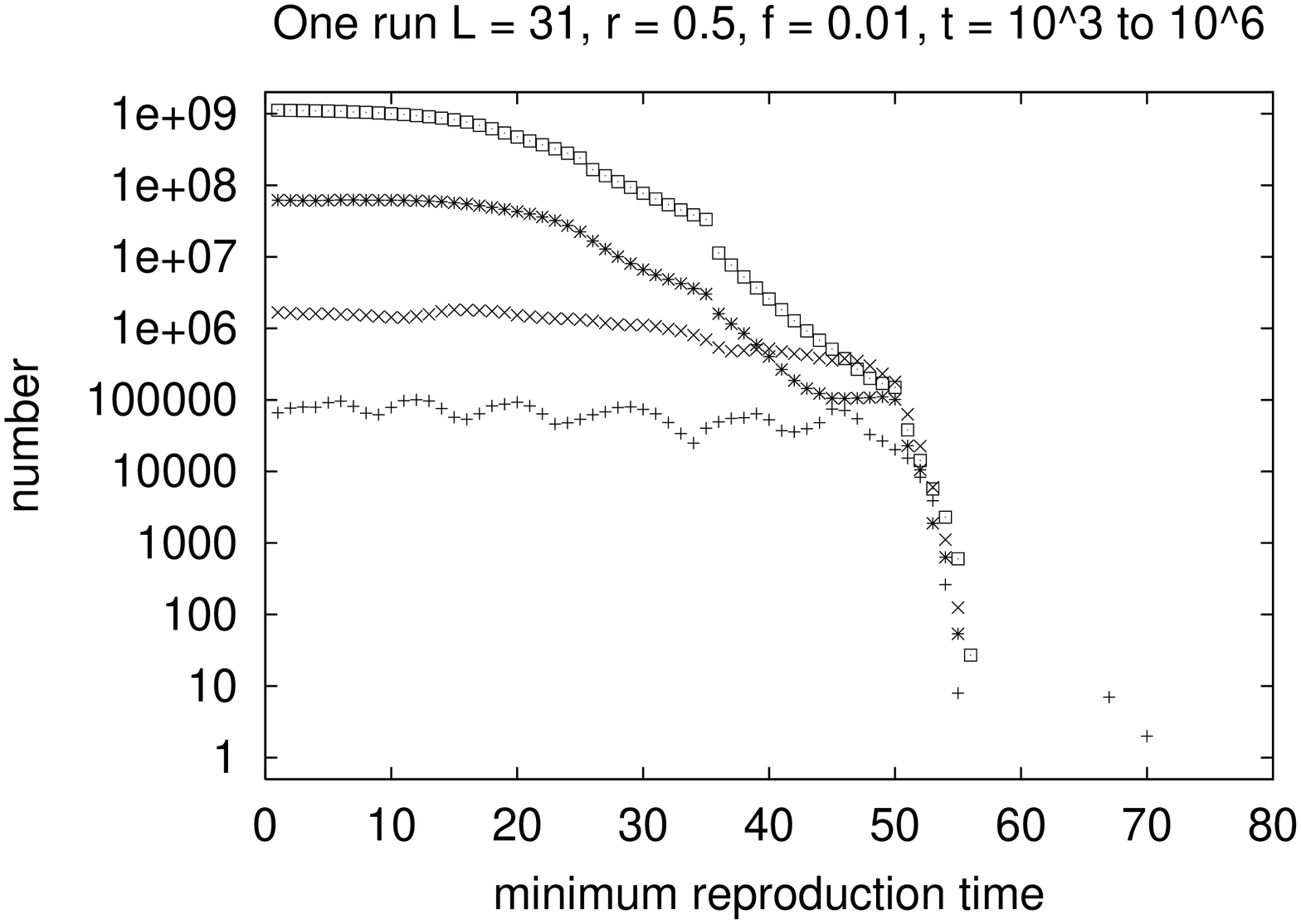}
\end{center}
\caption{With diffusion on a $31 \times 31$ square lattice, $f = 0.01, \; r = 0.5$.}
\end{figure}

\begin{figure}[hbt]
\begin{center}
\includegraphics[angle=-90,scale=0.33]{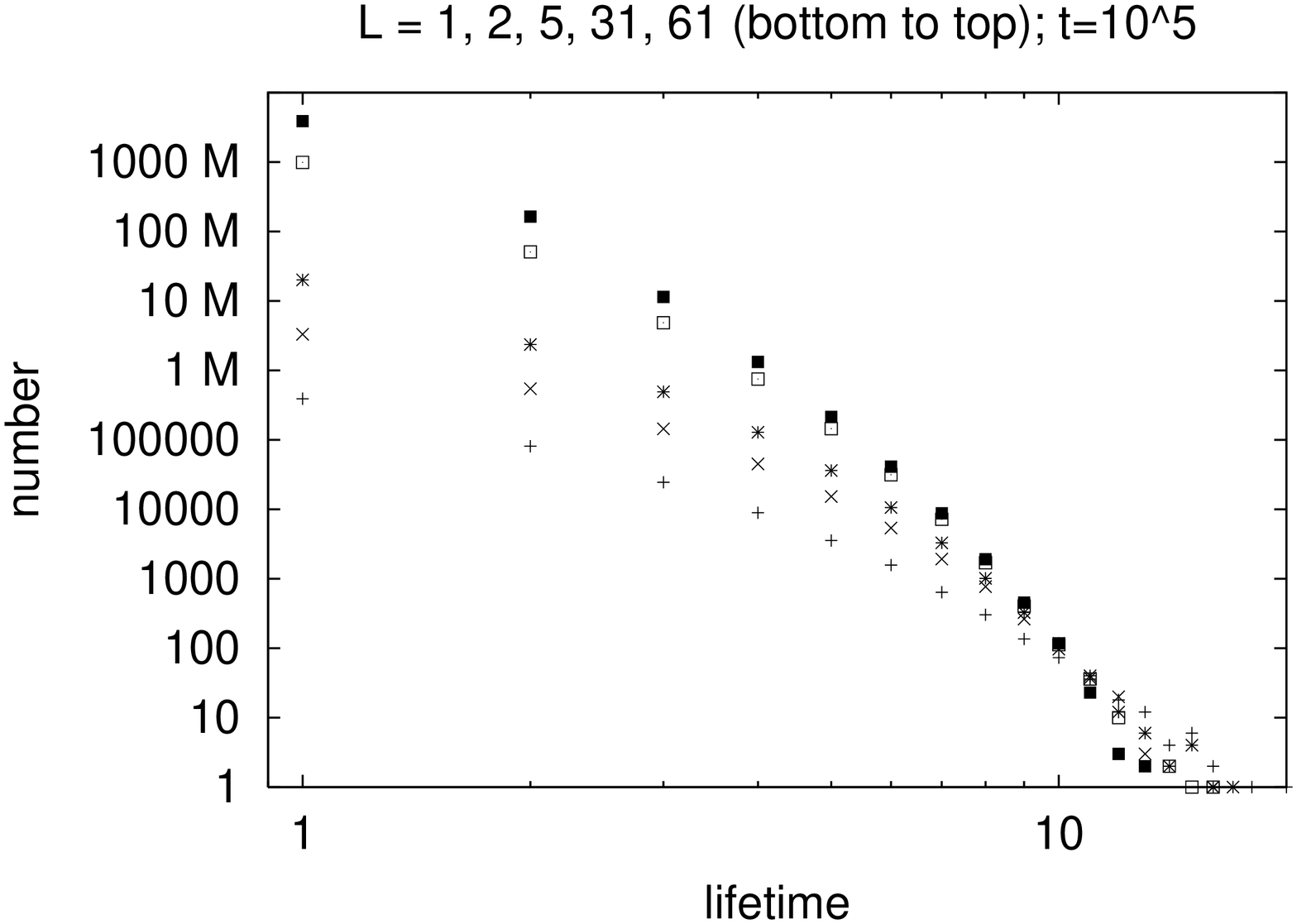}
\includegraphics[angle=-90,scale=0.33]{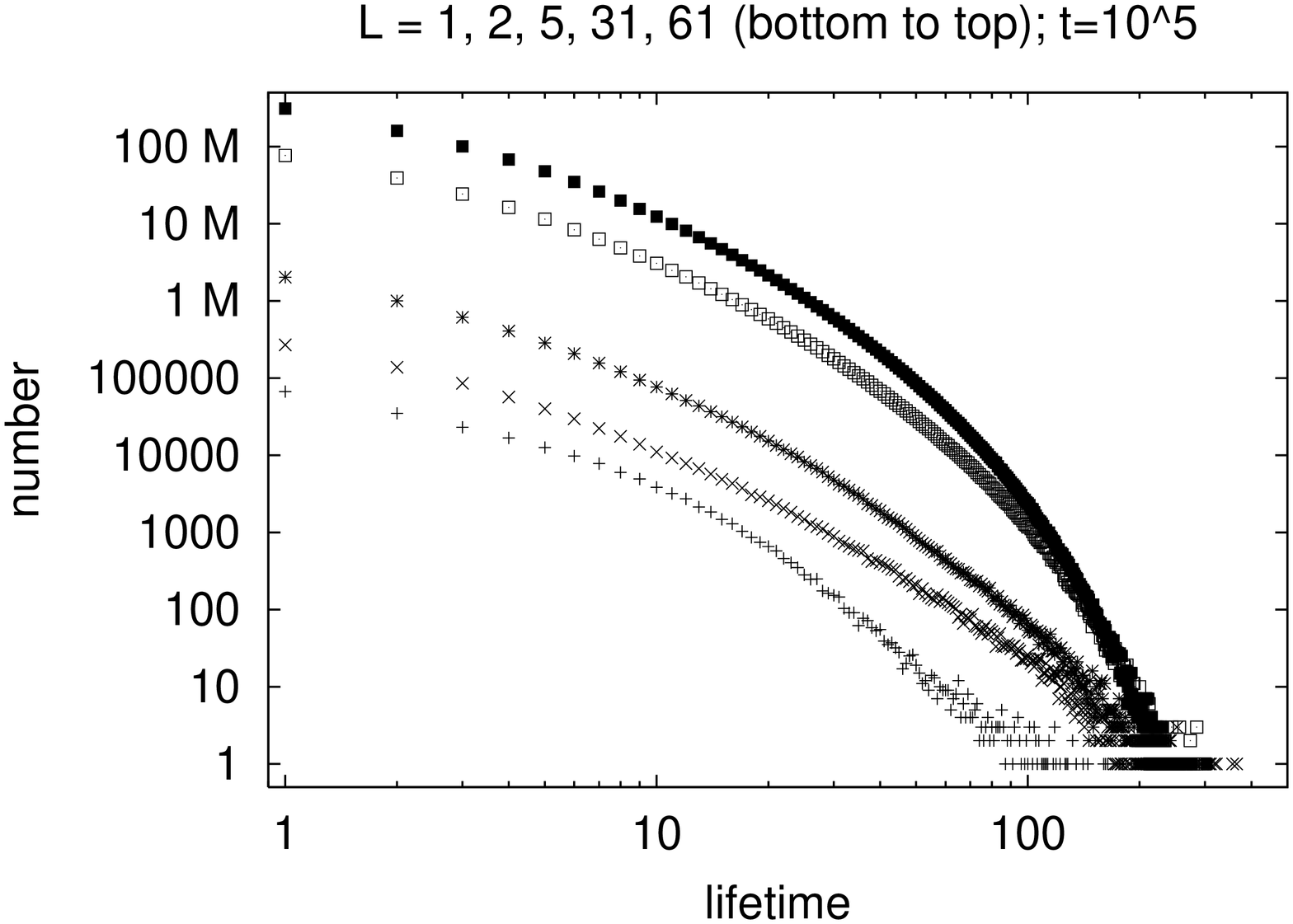}
\end{center}
\caption{Parameters as in Fig. 2 (Fig.6a, $r = 0.05$, without fitness test), 
and 4 (Fig.6b, $r = 0.005, f = 0.1$) for various lattice
sizes $L$ at $ t = 10^5$. Since only one
run was made, the resulting histograms increase with increasing $L$.}
\end{figure}

\begin{figure}[hbt]
\begin{center}
\includegraphics[angle=-90,scale=0.50]{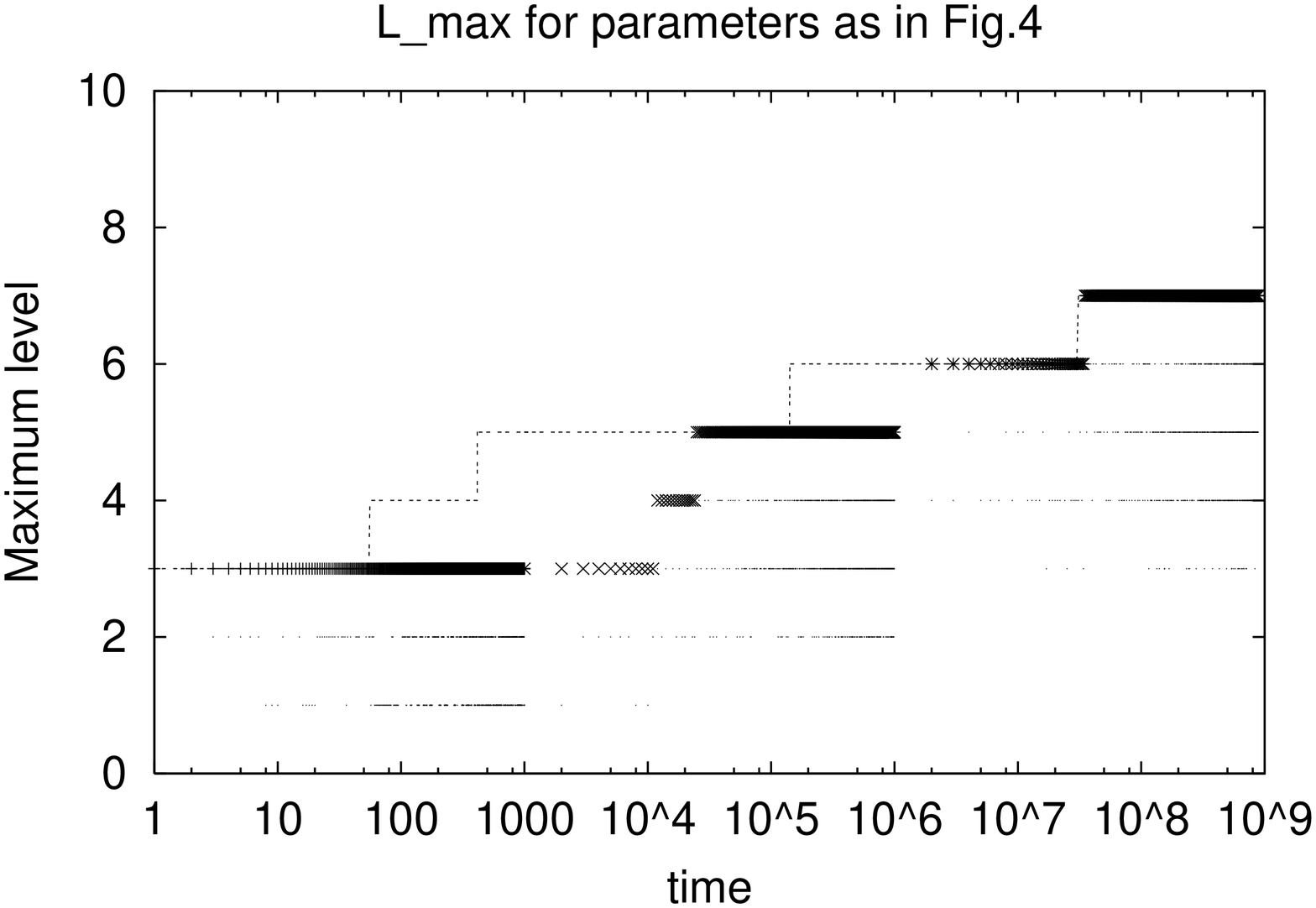}
\end{center}
\caption{Variation of $\ell_{max}$ with time up to nearly $10^9$ time steps
for the parameters of Fig.4, showing the less than logarithmic increase
at long times (big symbols). The small dots below the big symbols are the
actual numbers of occupied levels in the same simulations. The higher
dashed line gives $\ell_{max}$ for ten times higher probabilities of mutation 
and level formation. 
}
\end{figure}

\subsection{Variants}

In the above simulations, the lowest level is by construction always fully
occupied, but only a few species exist in the higher levels. We avoided 
the extinction of so many species by reducing the "birth punishment" factor
$r$ from 0.05 to 0.005, and by making the fitness test only with a probability
of $f$, that means, on average only at every $(1/f)$-th time step the
less fit species may go extinct. Now figure 4 for $f = 0.1$ shows for the 
lifetime distribution a more complicated behaviour, similar to
\cite{chow1,chow2,chow3,chow4}, without a lattice. For diffusion on a lattice,
the lifetimes may become log-normally distributed for these parameters. Fig.5
shows a more interesting distribution for $f = 0.01, \;r = 0.5$.  Simulations with
different system size give similar results, Fig. 6. In real palaeontology,
of course, it would be difficult to measure lifetime distributions over ten
decades with a single criterion, as is done here and in Fig. 2. Fig.7
suggests that the maximum number of layers increases very weakly; again 
times varying over nine decades are difficult to observe palaeontologially.

\section{Inhomogeneous eco-system: a individual-based description}

In this section we report results of the most detailed description
of the eco-system. We not only incorporate explicitly the birth,
ageing and death of individual organisms, but also take into account
the spatial inhomogeneities of the populations in terms of a lattice
model that is very similar to the one considered in the preceding
section.

Thus we assumed that each individual either ages by one time unit for 
each time step, or it dies. Besides the deaths as prey, unchanged from 
the previous section, the above selection according to fitness and 
litter size is replaced by a Gompertz-type death probability as a
function of the age $a$ of the individual, and depending also on the 
maximum age $X_{max}$ and litter size $M$ of the whole species:
$$p_{death} = \exp[(a-X_{max})r/M]$$
where $r$ is the same free parameter, e.g. 0.05, as in the earlier 
sections. (For ages below the minimum reproduction age $X_{rep}$ the
death probability is assumed to be age-independent, with $a$ replaced
by $X_{rep}$ in the above equation.) 

The survivors give birth to $M$ offspring at each time step with 
probability 
$$p_{birth} = (X_{max}-a)/(X_{max}-X_{rep})$$
if their age $a$ is above the minimum reproduction age $X_{rep}$. 
The just-born babies die with a Verhulst probability \cite{cebrat}
$1 - n/n_{max}$ where $n$ is the actual current population of the 
given species on the given lattice site, and $n_{max}$, taken here as 100,
is its maximum possible population. 

A population may expand into a neighbouring lattice site, if the population
there is zero for the same species on the same level. The population stays the
same at the old site, whereas at the new site, as in our standard sympatric
speciation assumption, the population is taken randomly between one and the 
population at the old site; also the age of the emigrants starts at zero.

\begin{figure}[hbt]
\begin{center}

\includegraphics[angle=-90,scale=0.33]{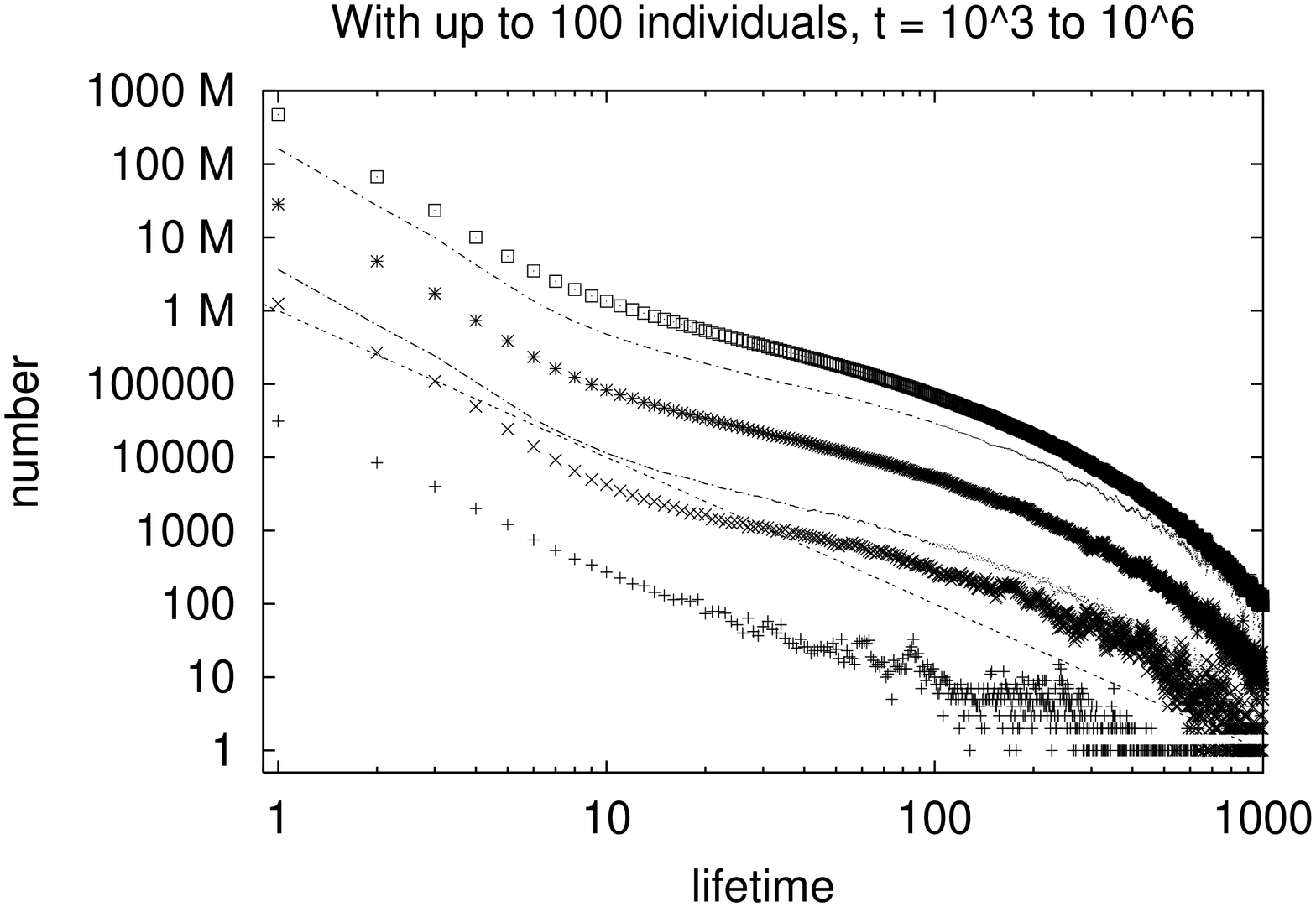}
\includegraphics[angle=-90,scale=0.33]{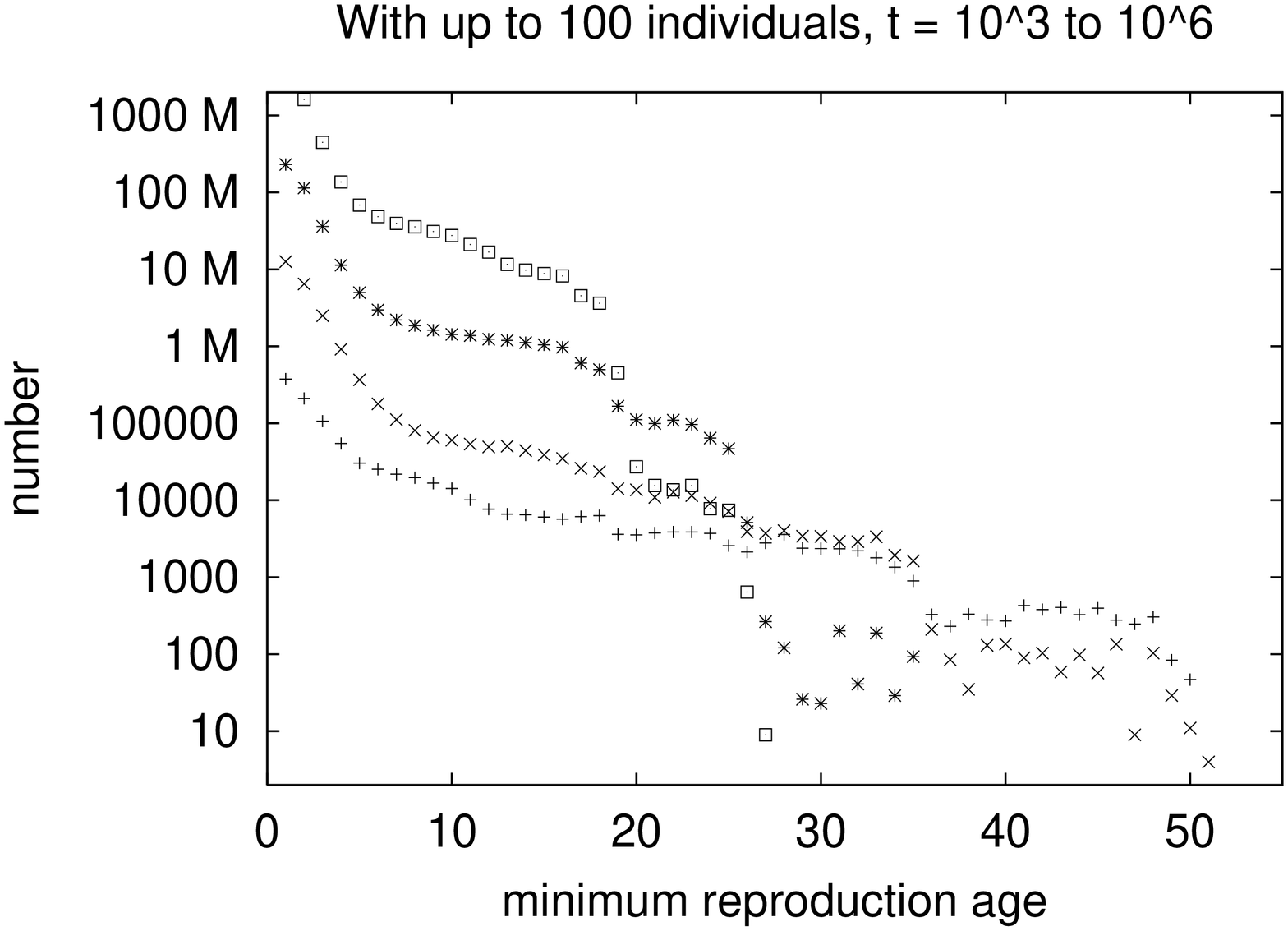}
\end{center}
\caption{
Distribution of lifetimes and minimum reproduction ages with individuals on
a $13 \time 13$ lattice at $t = 10^3, \; 10^4, \; 10^5 \; 10^6$. The thin
curves (or dots) refer to $t =10^5$ for $5 \times 5$ (lower data) and $31 
\times 31$ (higher data) for the lifetimes. 
}
\end{figure}

Fig.8 shows lifetimes and minimum age of reproduction with the individuals
on a $31 \times 31$ lattice; smaller lattices gave similar curves. To speed 
up equilibration, $p_{mut} = 0.01, \; p_{lev} = 0.001$ were taken ten times
higher than before. To prevent the number of levels to increase towards 10,
a new level could be created only if the total population on that lattice
site was at most $2 n_{max} = 200$. We see that roughly but not precisely 
the lifetimes follow an inverse-squared power law (straight line in Fig.8a).

\section{Discussion}

We have extended our earlier ``unified'' models of evolutionary ecology 
\cite{chow1,chow2,chow3,chow4} by taking into account the spatial 
inhomogeneities arising from migration of populations  from one patch to 
another. Besides, individuals and species, we also allowed evolution 
of genera (i.e., sets of species with the same food habits). 
By carrying out extensive computer simulations, we find that, 
depending on the details of the parameter set, the distribution of the 
lifetimes of the species may be simple exponential or, as before, a 
combination of power laws. Other aspects like, for example, the 
self-organization of the minimum reproduction ages, are less affected by 
this generalization.

\bigskip
We thank DFG/BMZ for supporting this collaboration through grant DFG/Sta130.

\end{document}